\documentclass[traditabstract]{aa}
\usepackage{natbib}
\usepackage{txfonts}
\usepackage{graphicx}

\titlerunning{A bow shock surrounding V CVn?}
\authorrunning{Neilson et~al.}

\begin{document}

\title{Evidence of a Mira-like tail and bow shock about the semi-regular variable V CVn from four decades of polarization measurements}

\author{Hilding R. Neilson \inst{1} \and Richard Ignace \inst{1}  \and  Beverly J.~Smith \inst{1} \and Gary Henson\inst{1} \and Alyssa M. Adams\inst{1,2} }
\institute{
   Department of Physics \& Astronomy, East Tennessee State University, Box 70652, Johnson City, TN 37614 USA
   \email{neilsonh@etsu.edu} 
 \and
 Department of Physics, Arizona State University, Tempe, AZ, 85287
  }

\date{}

\abstract{
Polarization is a powerful tool for understanding stellar atmospheres and circumstellar environments.  Mira and semi-regular variable stars  have been observed for decades and some are known to be polarimetrically variable, however, the semi-regular variable V Canes Venatici displays an unusually  large, unexplained amount of polarization.  We present ten years of optical polarization observations obtained with the HPOL instrument, supplemented by published observations spanning a total interval of about forty years for V~CVn.  We find that V~CVn shows  large polarization variations ranging from $1$ -- $6\%$.  We also find that for the past forty years the position angle measured for V~CVn has been virtually constant suggesting a long-term, stable, asymmetric structure about the star.  We suggest that this asymmetry is caused by the presence of a stellar wind bow shock and tail, consistent with the star's large space velocity.

}
\maketitle

\section{Introduction}

% Introductory paragraph - important of Miras and SRAs, future of the Sun, standard candles
Mira and semi-regular variable stars represent an important phase of stellar evolution as a future stage of evolution for our Sun and as bright standard candles.  They have been continuously observed for centuries \citep{Hatch2012}, but still present many mysteries. One example is the prototype Mira itself:  $o$~Ceti displays a far-ultraviolet  \citep{Martin2007} and H{\sc i} tail \citep{Matthews2008}, caused by its motion through the surrounding interstellar medium (ISM). Additionally, most Galactic Miras and semi-regular variable stars have measured period changes \citep{Uttenthaler2011} and exhibit polarimetric variability \citep{Serkowski2001}.
Period changes ultimately probe stellar properties of the interior and stellar evolution, while GALEX UV observations trace the long-term interaction between the stellar wind and the ISM. Polarization observations explore the space in between: the stellar atmosphere and circumstellar medium \citep{Clarke2010}.  In this work, we present forty years of polarimetry measurements for the M4e-M6IIIA:e semi-regular variable star V~Canes Venatici (V CVn, HD 115898; IRAS 13172+4547; HIP 65006) and discuss its anomalous behavior.

 The star V~CVn has a measured pulsation period of about 192 days and varies from about 8.4 to 7.1 magnitudes at optical wavelengths.  There have been a number of studies regarding its pulsation mode. \cite{Kiss2000} suggested that the star displays two pulsation periods, 194 and 186 days respectively, and that the pulsation shows a stable biperiodic variation. \cite{Buchler2004} also found evidence for resonance between two pulsation modes in a sample of semi-regular variable stars including V~CVn. \cite{Bedding2003} suggested that semi-regular variable stars undergo solar-like oscillations, that are stochastically excited by convective motions. 

% Polarimetric observations of Miras
A number of polarization studies have been presented for a plethora of Mira variable stars and smaller amplitude semi-regular variables.  \cite{Shakhovskoi1963} presented the first detection of variable polarization in a Mira, while independent observations of a number of Mira variables and semi-regular variable stars, including the prototype $o$~Ceti, suggest that intrinsic polarization is a ubiquitous property \citep{Serkowski1966}.  \cite{Serkowski1966} noted that V~CVn showed the greatest amount of polarization, with $p\approx 6\%$.  

% Polarimetric variability
\cite{Serkowski1966} also found variable polarization and a near-constant position angle for V~CVn. The star has maximum polarization at minimum brightness and minimum polarization at maximum brightness, at least over the one pulsation cycle that was observed. While the authors suggested a number of possible scenarios to explain this intrinsic polarization measured in the Mira variable stars, none of them account for the anomalous polarization of V~CVn nor for that seen in another semi-regular variable L$_2$ Puppis.    \cite{Dyck1971} reported near-infrared JHK polarization measurements of V~CVn and found $p < 1\%$ and that the polarization and position angle both decrease as a function of wavelength.  More recently, \cite{Serkowski2001} presented optical polarimetry for 167 cool stars, including V~CVn, where the position angle varied fro about $99^\circ$ to $122^\circ$ and the polarization varied from about 1\% to 8\%.  Polarization has also been detected in samples of Carbon stars \citep{Lopez2011}, asymptotic giant branch stars, and R CrB stars \citep{Bieging2006}. These results suggest that most if not all cool stars have some intrinsic polarization, but only two, V~CVn and L$_2$~Pup appear to have large polarization. \cite{Poliakova1981} and \cite{Magalhaes1986a} presented observations of V~CVn.   \cite{Magalhaes1986b} also described polarization seen in L$_2$ Pup that was as large as $8\%$ and varied with pulsation phase.  \cite{Magalhaes1986a, Magalhaes1986b} also found that the position angle appears to vary, from about $100^\circ$ to $120^\circ$ for V~CVn and from $150^\circ$ to $180^\circ$ for L$_2$ Pup.

Polarization has also been detected in the Balmer lines of Miras \citep{Mclean1978}; however, that polarization may be caused by a different mechanism than the polarization observed in the continuum. The Balmer line may relate to shocks generated by pulsation \citep{Bowen1988}, which could affect polarization in the continuum as a function of pulsation phase.  Similar polarization was measured for other lines in $o$~Ceti \citep{Tomaszewski1980}. The possibility that shocks are the source of polarization in the lines was explored by \cite{Svatos1980}, who observed a number of Mira variable stars. He found that near the pulsation phase $\phi = 0.8$ in the brightness curve the  polarization appeared to peak. \cite{Svatos1980} argue that this polarization peak is caused by UV-irradiated silicate grains, whereas the UV flux is due to shocks propagating in the photosphere.  \cite{Magalhaes1986a} measured line polarization in TiO 4955~$\AA$ and CaI 4226~$\AA$ and also found significant polarization of up to 6\%. 

  In spite of all of the observations of V~CVn, the high polarization fraction spanning many pulsation cycles is still not understood.    In this work, we present forty years of polarization measurements for V~CVn and hypothesize that the variable polarization along with a nearly constant polarization position angle can be explained by the presence of a debris disk or by a bow shock and Mira-like tail. We discuss the observations in Sect.~2.  In Sect.~3, we suggest and explore potential scenarios to explain these observations and in Sect.~4 we summarize our results.

% Outline of remainder of paper

\section{Polarization Observations}
\begin{figure}[t]
\begin{center}
\includegraphics[width=0.5\textwidth]{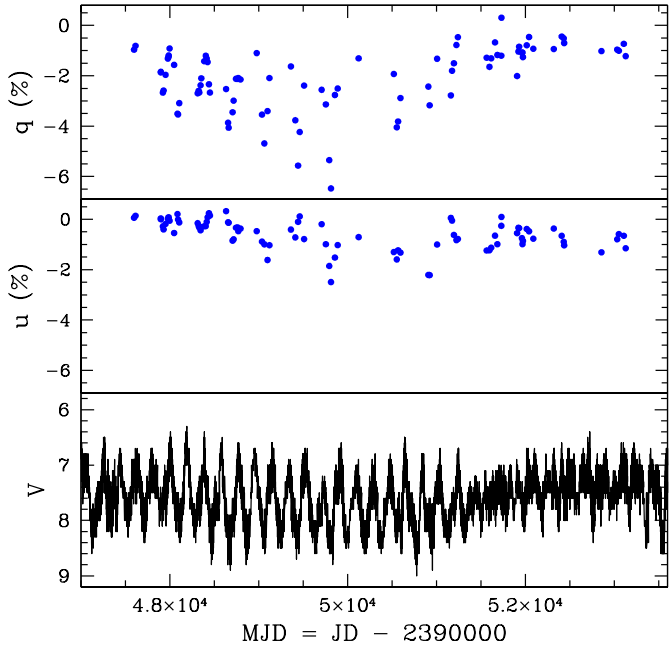}
%fqu.eps}
\end{center}
\caption{Measured HPOL $Q$ (top panel) and $U$ (middle panel) stokes parameters as a function of time for V~CVn along with the corresponding V-band light curve (bottom panel).}\label{fqu}
\end{figure}

% Present data and time span, mention ISM issues
Several Mira variable stars were observed using the HPOL spectropolarimetric instrument from 1990 - 2002 \citep{Wolff1996}.  HPOL simulates broadband photopolarimetry by integrating the spectropolarimetric measures with bandpass filter functions for Johnson BVRI.  Details can be found at archive.stsci.edu/hpol.  Here we consider $Q$ and $U$ Stokes fluxes for V~CVn that have been observed regularly with HPOL and supplement these with measurements presented in the literature by \cite{Poliakova1981} and \cite{Magalhaes1986a}.  In Fig.~\ref{fqu}, we present the measured $Q$ and $U$ stokes parameters, as well as the optical light curve for a time span ranging from 1965 to 2002, almost forty years.  The visual observations are from the AAVSO\footnote{The AAVSO refers to the American Association of Variable Star Observers; see www.aavso.org.}. 

% Present Q & U stokes parameters as a function of time
The $|Q|$ and $|U|$ Stokes data vary from nearly 0\% up to about 7\% and nearly 0 to $3\%$, respectively.  The two appear to show similar behaviors as a function of time.  There is a hint that the amount of variability in $Q$ and $U$ decreases when the $V$-band amplitude is smaller, suggesting that the source polarization correlates with the brightness.

\begin{figure}[t]
\begin{center}
\includegraphics[width=0.5\textwidth]{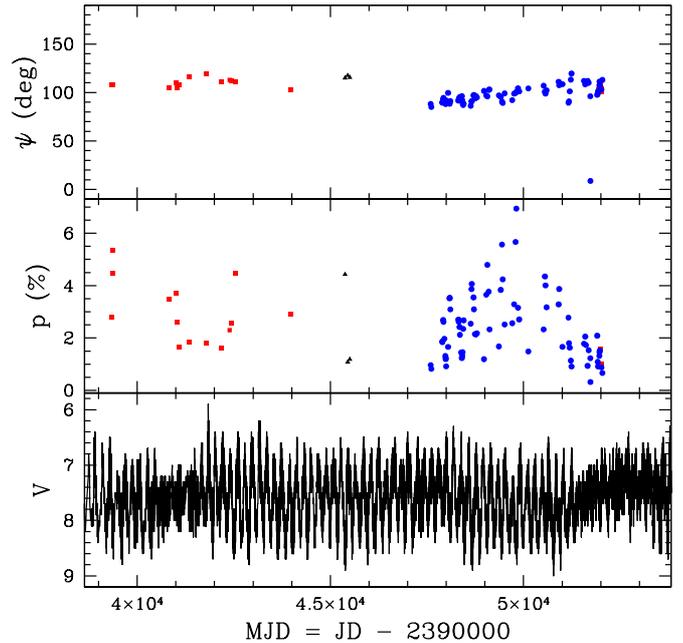}
\end{center}
\caption{Measured position angle, $\psi$, in degrees; polarization; and visual brightness for V~CVn from the AAVSO database, where the red squares are data from \cite{Poliakova1981}, black triangles are from \cite{Magalhaes1986a} and the blue circles represent the HPOL data.  Differences between this plot and Fig.~\ref{fqu} are because of  the additional time span from MJD $= 3.9\times 10^4$ -- $4.4\times 10^4$ considered. The corresponding V-band light curve is included for comparison. }\label{f1}
\end{figure}

% Present p and PA as a function of time
We have transformed the measured HPOL $Q$ and $U$ Stokes parameters to polarization, $p$, and position angle $\psi$ for our analysis.   The polarization is defined as $p^2 = Q^2 - U^2$ while the position angle is $\phi = \arctan(U/Q)$. In Fig.~\ref{f1}, we present $p$ and $\psi$ along with the optical light curve for comparison.  Note that the measured polarization is not corrected for the polarizing influence of the interstellar medium; however, intrinsic source polarization is made evident via variability, since any interstellar contribution is constant.  The polarization is seen to vary from about 2\% to 6\% for V~CVn, a factor of $3\times$.  By contrast the position angle $\psi$ varies over only a restricted range of $100^\circ$ to $120^\circ$ over forty years. 

 Linear polarization is a valuable diagnostic of the geometry of a star and circumstellar regions. A centro-symmetric pattern of polarization produces no net polarization for an unresolved source.  The variable polarization seen for V~CVn signifies a non-spherical geometry.  The nearly constant position angle signifies that the geometry is long-term stable. As a general rule, stochastic variations in geometry, such as wind clumping, would yield variable polarization and random variations in $\psi$.  The presence of binarity is normally revealed by smoothly varying polarization and position angle \citep[e.g.][]{Brown1978}.  A constant polarization position angle is suggestive of a fixed but non-spherical geometry, such as a circumstellar disk or bipolar jet \citep{Brown1977}; then variable polarization could arise from a changing level of optical depth (either opacity, density, or both).

% Note apparent rise in PA
While $\psi$ is approximately constant, there may be a hint of a gradual increase in the HPOL data.  We compute the slope of  $\psi$ as a function of time assuming a linear function.  Using only the HPOL data, the slope is  consistent with an increase from about $\psi = 90^\circ$ to $120^\circ$ over the time span of about a decade.  However, when we fit the entire data set, the best-fit line is consistent with zero slope.  It is not obvious whether the gradual position angle increase is a real phenomenon or if there are correlated errors in the HPOL data.  

% Note box-like behavior in p
As mentioned, the polarization measurements are not corrected for interstellar polarization.  Because of the observed variability then the ISM contribution must be at most the minimum polarization.  The minimum polarization is about $1\%$, and the star is located at $70^\circ$ Galactic latitude at $\approx 700~$pc \citep{Serkowski1966, Leeuwen2007}.  Based on the results of \cite{Materne1976}, the contribution of ISM polarization is small at that distance. Therefore, it is expected that interstellar polarization will not contribute significantly to the measured polarization of V~CVn, almost at a level less than $1\%$.

\begin{figure}[t]
\begin{center}
\includegraphics[width=0.5\textwidth]{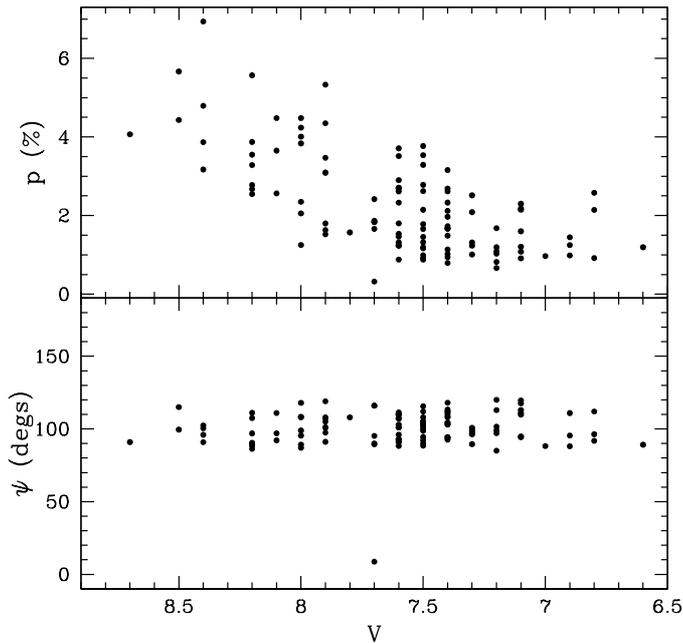}
\end{center}
\caption{Measured polarization percentage; and position angle, $\psi$, in degrees for  V~CVn (Right) from our dataset as a function of $V$-band brightness from the AAVSO database.}\label{f2}
\end{figure}

The HPOL polarization data also display a ``box-like'' pattern.  Polarization is smallest at the beginning and end of the HPOL data, but peaks near Julian date $MJD\sim 50000$.  This behavior could indicate variability over time scales greater than the pulsation period, or it could be coincidental.  The HPOL polarization data were accumulated over more than twenty pulsation cycles, hence the box-like structure could simply arise from sparse observations over different cycles, where the observations miss the peak polarization during pulsation cycles at the beginning and end of the HPOL data.  We check this by plotting polarization and position angle as a function of the AAVSO visual brightness in Fig~\ref{f2}.

% Describe p and PA as function of V
Plotting polarization and position angle as a function of $V$-band brightness is similar to folding the data into one pulsation cycle.  Folding the data is less than ideal as the amplitude is variable and the pulsation period is not precisely measured.  Still, comparing polarization and the $V$-band data indicates an apparent inverse correlation, in which maximum polarization occurs at flux minimum and minimum polarization occurs at flux maximum.  We also tested the data by checking the polarization during flux increase and decrease to explore any potential phase lag or lead. However, there was no direct correlation to indicate any such phase effect.

% Note test for any cyclic behavior and lead and lag time, however, data is insufficient for 
% Test PA data for slope, over all slope is consistent with 0
While there is a correlation between polarization and $V$-band flux, there is no correlation between the visual flux and the position angle.  Fitting a simple linear relation between $V$ and $\psi$, we find the slope is again consistent with zero, suggesting that the position angle is constant within the scatter of the data.  

% The data describes two phenomena: p is inversely aligned with V and PA is nearly constant with possible slow increase.
The HPOL observations provide new perspectives on the structure of V~CVn and semi-regular variable  stars in general.  The polarigenic mechanism in this star is still a mystery and any model should not only explain the polarization variability, but also clarify
\begin{enumerate}
\item the inverse correlation between the visual light curve and the polarization and 
\item the near-constant position angle as a function of time over the past forty years.
\end{enumerate}
In the next section, we outline a number of scenarios that might explain these polarization observations and discuss their viability.

\section{Asymmetric Structure of V~CVn}
Based on the polarization observations, we find that there must be a long-term stable asymmetry to produce the fixed position angle for V~CVn, as inferred from Fig.~\ref{f2}. 
Potential causes include star spots, a binary companion, rotational distortion of the star, a circumstellar disk, or a bow shock and Mira-like tail. However, based on previous observations of Mira and numerous other cool variable stars \citep{Serkowski2001}, it is reasonable to assume that V~CVn exhibits at least some intrinsic polarization due to pulsation-driven shocks \citep{Fabas2011} and/or circumstellar dust \citep{Ireland2005} as suggested by models.  Consequently, some component of the observed polarization may arise from these effects; however, for V CVn, the polarimetric amplitude is quite large, and the fixed position angle is atypical, suggesting another contributing mechanism.

\subsection{Rotation}
One possible mechanism to generate a non-spherically symmetric stellar wind is rapid rotation.  Rapid rotation both distorts the effective temperature and gravity of the star \citep{Ziepel1924} and potentially leads to a centrifugally-driven wind.

While rapid rotation could explain the nearly constant position angle, it would raise a number of challenges as well. One such issue is that rapid rotation would significantly alter how the star pulsates. Rotation acts to split non-radial pulsation frequencies and damp radial pulsation. If V~CVn was rotating near critical velocity, then it would be difficult to understand how it pulsates radially \citep{Chandrasekhar1968}. Furthermore, it is also difficult to understand how a star can evolve to such a late stage and rotate at such a high rate, near critical velocity.  Mass loss at earlier stages of evolution decreases the star's angular momentum and decreases the rate of rotation.  The only way for V~CVn to spin near critical velocity is to accrete significant material from a companion \citep{Tout2012}.  Such a close companion would likely have been detected (see Sect.~3.3) or have merged with V~CVn. If V~CVn was the product of a merger and rapidly rotating then the merger would have been relatively recent and V~CVn should appear younger and have significant chemical anomalies \citep{Perets2009}.  There is currently no evidence to support such scenarios, hence we argue that rapid rotation is very unlikely.  For instance, polarization measurements of rapidly-rotating Be~stars range from less than one percent to a few percent \citep[e.g.][]{Wheelwright2012},  meaning that V~CVn would also have to rotate at nearly critical velocity to explain the large polarization.  That rotation, however, will not explain the polarization variability.

\subsection{Spot Model}
A second potential explanation for the polarization is star spots.  Star spots lead to a stellar radiation field that is not circularly symmetric, which can lead to a non-random polarization position angle at a given time \citep{Al1992}.  These spots may be bright spots caused by giant convective cells like those observed in Betelgeuse \citep{Haubois2009} or by magnetic field interactions, like those observed in the Sun or in RS~CVn stars \citep{Muneer2010}.  However, star spots are problematic for explaining the position angle of V~CVn because $\psi$ is nearly constant for about four decades.  One spot on the stellar surface would shift location as the star rotates, unless the rotation period were of the order of a millennium. Also, convective cell and magnetic field spot time scales are not long enough to maintain a static structure for forty years. This suggests that to maintain a constant position angle for forty years, there must be a continuous band of spots across the stellar surface. But if spots are located at random areas on the stellar surface, then the position angle would tend to vary in a random way.  Actually, random variations in position angle are observed in other Mira stars, suggesting that spots could be a viable explanation for their behavior.  

Although spots might contribute to the polarization of V~CVn, they cannot account for the long-term fixed position angle.  There is no reasonable mechanism for star spots to appear coherently in a band on its stellar surface. Convective cells would seem to brighten and dim at random locations across the stellar surface while magnetic spots would drift.  It is thus unlikely that the polarization of V~CVn is governed by starspot activity.

\subsection{Binary Companion}
Binarity is another potential solution.  A stellar companion would contribute to the polarization fraction as an asymmetric source of radiation interacting with the dusty wind of V~CVn.  However, to cause a constant position angle as a function of time, the orbital period must be sufficiently greater than the observed time scale of the position angle. Following the analytic derivations by \cite{Brown1977} and \cite{Brown1978}, we calculate the position angle as a function of orbital phase and inclination. For a circular, edge-on orbit, consistent with the minimum possible orbital period, the predicted position angle varies by $10^\circ$ for the orbital phases $\phi = 225^\circ$ -- $270^\circ$.  If this range of $\phi$ corresponds to the past forty years of observations then the minimum orbital period is about 320 years. We ignore highly eccentric orbits because tides would circularize the binary orbit during earlier red giant stages of evolution \citep{Hut1981}. Assuming that V~CVn and its hypothesized companion have a total mass of about $4~M_\odot$, then the minimum orbital separation is about $2\times 10^4~R_\odot$.

 While a sufficiently large orbital separation and orbital phase range is consistent with a constant position angle, the given orbital phases correspond to a minimum polarization fraction assuming the optical depth of the circumstellar material of V~CVn and stellar flux is constant.  However, the polarization fraction is proportional to $\tau/[1 + L(t)/L_2]$, where $\tau$ is the optical depth, $L(t)$ is the flux of V~CVn and $L_2$ is the companion's flux \citep{Brown1978}.  Based on this relation, the predicted polarization is consistent with the observed polarization fraction if  the companion star is significantly brighter than V~CVn.  In that scenario $p \propto \tau$, and $\tau$ must vary as a function of pulsation phase by a factor of 2 -- 3. However, the brighter companion would have been detected.  If the companion is much less luminous than V CVn, then the binary system will appear roughly symmetric and the polarization much smaller than that observed. Therefore, a binary companion is an implausible explanation for the observed polarization.  We also note that \cite{Famaey2009} found no evidence that V~CVn is a spectroscopic binary.

\begin{figure*}[t]
\begin{center}
\includegraphics[width=0.9\textwidth]{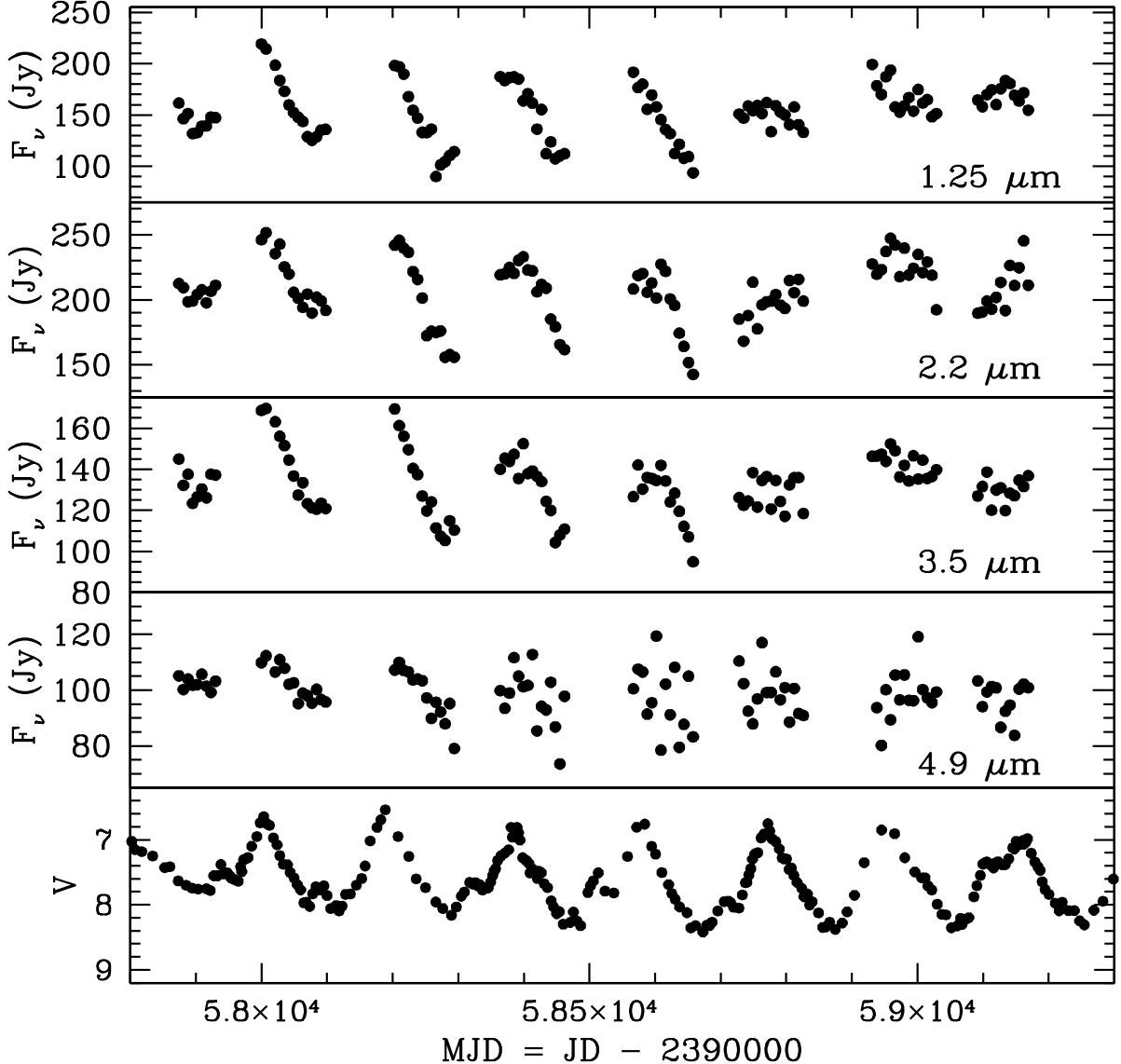}
\end{center}
\caption{Infrared light curves for V~CVn from COBE observations as a function of time \citep{Price2010}. }\label{fir}
\end{figure*}

% must be most oblate at max p and most circular min p or min L and max L respectively
\subsection{A Dusty Disk}
A circumstellar disk would lead to a constant position angle as a function of time and would also contribute to the polarization fraction as well.  To explain the observed variation in Fig.~\ref{f2}, the disk would have to be about one-half as bright as V~CVn itself when at brightness minimum at optical wavelengths.  A dusty disk, however, is typically cool with temperatures less than $1500~K$, hence it would not be sufficiently bright at these wavelengths to cause the observed polarization itself.

% Oblateness argument
We suggested that rotationally induced oblateness cannot explain the observed polarization variability nor maximum polarization simply because the star cannot be oblate enough.  However, if the star hosts a dusty disk that scatters radiation, then we can consider maximizing the polarization based on \cite{Brown1977}, for which the geometric factor becomes $\gamma = 0$ for the relation $p = \tau (1-3\gamma)$.  This situation can easily lead to a polarization, $p = 6\%$, analogous to Be star disks \citep[e.g.][]{Hiltner1956, Brown1989, Haubois2014}.  However, it is not clear that a disk can explain the polarization variability.  

One possibility is that the polarization is linked to radius variations. The seminal work of \cite{Brown1977} adopted a point source approximation in its evaluation for circumstellar polarizations.  \cite{Cassinelli1987} derived a correction for the influence of a star's finite size.  This correction results in generally lowering the polarization, since the radiation field near the star is less anisotropic (in other words, a point source always maximizes the anistropy of the radiation field).  Consequently, for a fixed circumstellar structure, a star that changes in size ultimately modulates the polarization by virtue of a relative change in the anisotropy of the radiation field with pulsation phase.  However, we can rule out the possibility that the polarization is linked to radius variations because the flux and radius variations have a significant phase lag, hence would violate the measured correlation between flux and polarization.

Another possibility is a change in the disk optical depth.  For optically thin scattering, the polarization is linear in optical depth, with $p \propto \tau$.  Mid-infrared IRAS spectra for V~CVn displays a strong silicon emission feature at 10~$\mu$m \citep{Olnon1998,Sloan1998}.  This emission feature suggests that the circumstellar medium about V~CVn is not significantly dense  to be optically thick \citep{Kyung2011}.  A star moving rapidly through the ISM would not be able to build up a dense circumstellar shell, hence would not become optically thick at 10~$\mu m$.
With variations of $p$ from 1\% to 6\%, the disk opacity, density, or both would need to change by a factor of six to explain these variations.  The optical depth of the disk must be at maximum at flux minimum and minimum at flux maximum and might be a result of variations in the dusty wind that interacts with the disk.  At minimum light, which corresponds to minimum effective temperature, more dust may form at the edge of the disk to increase the optical depth and polarization.  As the star brightens, the dust is accelerated away from the star and the disk, decreasing the disk optical depth again.  This is consistent with observations of the analogue star L$_2$~Pup  suggestive of episodic dust obscuration events \citep{Bedding2002}. However, this sort of stellar wind would be expected to interact with the disk and potentially disrupt it. 

This concept has physical plausibility, since we understand that the condensation radius generally scales inversely with temperature.  Additionally, the disk morphology is consistent with the observation of a fixed polarization position angle:  whether the dusty wind is stronger or weaker, the disk defines the overall departure from spherical symmetry as projected onto the sky.  The combination of the two phenomena can also explain why the polarization fraction for V~CVn is greater than what is typically observed for Mira variable stars and $o$~Ceti itself  \citep{Kruszewski1968, Serkowski2001}. 

Is there any other supporting evidence for the dusty disk hypothesis?  One might expect the presence of an IR excess from such a structure. We have examined COBE infrared flux measurements of the star \citep{Price2010}. Fig.~\ref{fir} displays the IR observations as a function of time along with the optical light curve.  While there are a number of gaps in the COBE data, the IR light curves appear roughly in phase with the optical light curve, which differs from Mira variable stars, but agrees with SRa stars \citep{Smith2003}.   However, time-averaged DIRBE infrared colors are consistent with observations of other semi-regular variable and Mira stars \citep{Smith2003, Smith2004}, hence we cannot draw conclusions from the above mentioned observations about the presence of a disk structure about V~CVn.

% To highlight the presence of IR excess, Fig.~\ref{see} shows the spectral energy distribution of V~CVn, including UBV Johnson data \citep{Ducati2002} and the COBE data \citep{Smith2004}. The IR fluxes are time-averaged with the vertical bars representing the amplitude of brightness variation (i.e., these are not error bars).  A blackbody function is overplotted for illustrative purposes; the curve is not a fit to the data.  The blackbody curves serves to illustrated the level of excess and the significant departure from a Rayleigh-Jeans slope. %\begin{figure}[t]
%\begin{center}
%\includegraphics[width=0.5\textwidth]{sed.eps}
%\end{center}
%\caption{Spectral energy distribution for V~CVn where optical data are from \cite{Ducati2002} while the infrared fluxes are from COBE data \citep{Smith2004}.  The accompanying blackbody function is not a best-fit but normalized to consistent with the flux at 1.25~$\mu$m.}\label{see}
%\end{figure}

\cite{Kervella2014} present new infrared images of L$_2$~Pup, which displays similar polarimetric behaviors as V~CVn, and detect a circumstellar disk that is approximately edge-on.  They also find evidence for circumstellar material beyond the disk, in the form of loop, bow-shock-like structures and geometry that is aligned in a north-east to south-west axis that is hypothesized to be an interaction between the stellar wind and an unseen companion.  The origin of this disk was not discussed but may be related to the formation of bi-polar planetary nebulae.  These observations are intriguing and motivate similar infrared observations for V~CVn.

\subsection{Bow shock and Mira-like tail}
% Disk can fit the  polarization but is awkward in explaining the p variation.  Requires strange interaction with the wind
A stellar disk about V~CVn explains the observed near-constancy of the position angle, but cannot easily explain the polarization variability.  We suggest that the variability is related to the interaction between a pulsation-dependent dust-driven wind and a debris disk, but it is not obvious how this would modulate the polarization.  Infrared excess in V~CVn may be qualitatively consistent with a debris disk, but does not stand out relative to observations of other semi-regular variable stars. 

One other explanation could be that the observed polarization and position angle are the result of a stellar wind bow shock and tail, analogous to that observed for Mira \citep{Martin2007, Matthews2008}.  Bow shocks and wakes have also been detected for a number of AGB stars \citep{Matthews2007, Matthews2011, Matthews2013, Libert2010}.

\begin{table}[t]
\caption{Velocities and polarization measurements for the entire sample of Mira and SRa stars from the HPOL database \citep{Wolff1996}.}
\label{t1}
\begin{center}
\begin{tabular}{lcccc}
\hline
Star & $\pi$ (mas) & $V_{\rm{rad}}$ (km~s$^{-1}$)& $V_{\rm{tan}} $ (km~s$^{-1}$) & p (\%)\\
\hline
V~CVn & $1.44\pm 0.86$ & $-4.70\pm 0.61$ & $133 \pm 82$ & 1 - 6 \\ 
$o$~Ceti & $10.91\pm 1.22$ & $63.5 \pm 0.6$ & $103\pm 12$ & 0 - 1 \\
Y Tau & $2.80\pm 1.01$ & $17.0\pm 4.4$ & $9\pm4$ & 0 - 1.5 \\
$\eta$ Gem & $8.48\pm 1.23$& $22.39\pm 0.36$&$36\pm 6$& $< 1$ \\ 
$\chi$ Cyg & $5.53\pm 1.10$ & $1.60\pm 0.60$ & $37\pm 8$&0 - 2\\
$o ^1$ Ori & $5.01\pm0.71$&$-8.40\pm0.23$ & $52\pm 8$ &$\approx 1$\\
R Lyr & $10.94\pm 0.12$&$-27.15\pm 0.20$ & $37\pm7$&$< 1$ \\
$\rho$ Per & $10.60\pm 0.25$ & $30.81\pm 0.11$ &$ 74\pm 2$&$< 1$ \\
U Hya & $4.80\pm0.23$&$-25.8\pm 1.7$&$ 56\pm 3$&$< 1$ \\
UU Aur & $0.38\pm0.41$ & $13.4\pm 0.8$ &$248\pm 272$ &$< 1$ \\
Y CVn & $3.12\pm 0.34$&$15.3\pm0.4$&$ 20\pm 3$&$< 1$ \\
g Her &$9.21\pm 0.18$&$1.49\pm 0.38$&$16\pm 0.4$ &$< 1$ \\
R Leo & $14.03\pm 2.65$&$14.1\pm 0.6$&$ 14\pm 3$&0 - 2\\
L$_2$ Pup&$15.61\pm 0.99$&$53.0\pm0.7$&$104\pm 7$& 1 - 6 \\ 
\hline
\end{tabular}
\note{Polarization measurements  for L$_2$ Pup from \cite{Magalhaes1986b}.}
\note{Values of parallax, radial and tangential velocities are from the Simbad database (http://simbad.u-strasbg.fr/simbad/) as reported by \cite{Leeuwen2007}.}
\end{center}
\end{table}

Polarization from bow shocks has been observed for a number of stars near the Galactic center \citep{Rauch2013, Buchholz2013}, where $p\approx 3$\%.  As a bow shock is an asymmetric wind structure, it could also explain the observed position angle.  A bow shock would in addition be consistent with the observed IR excess and the fact that V~CVn does not have an IR excess that differs from other Mira and semi-regular variable stars.  Further evidence supporting this idea is the measured proper motion and space velocity for V~CVn. The star appears to have a tangential velocity of about $133 \pm 82~$km~s$^{-1}$ \citep{Leeuwen2007}, similar to that of $o$~Ceti. The star also has a radial velocity of about $5~$km~s$^{-1}$, hence the space motion is almost completely tangential from the  perspective of Earth.  Velocities and HPOL polarizations are shown in Tab.~\ref{t1} for a sample of Mira and semi-regular variable stars. The motion is consistent with the presence of a bow shock and tail. Such a large space velocity might disrupt any sort of stellar disk. We list the velocities and polarization measurements for a sample of Mira and semi-regular variable stars observed using the HPOL instrument, along with that of L$_2$~Pup. There are four stars with significant space velocity, including both V~CVn and L$_2$~Pup, suggesting that a stellar wind bow shock is a promising solution to the question of polarization in V~CVn, especially if the bow shocks have greater density than observed for Mira.
% Bow shocks are another asymmetric structure that would yield a constant PA.  See Rauch, Neilson, etc.

% IR excess is consistent with other stars, so a debris disk must be optically-thin relative to the already optically-thin  wind

% High space velocities, include table

However, the presence of a bow shock analogous to that observed for the prototype $o$~Ceti raises two important questions. The first question is why would polarization be so significant for V~CVn and L$_2$~Pup but not for Mira?  The second question would be, why the polarization is variable.  Mira has a velocity similar to that of V~CVn and is much closer to Earth \citep{Martin2007}.  The bow shock and tail density are proportional to the number density of the local ISM and the stellar mass-loss rate, $\rho_{\rm{shock}} \propto (\dot{M}n_{\rm{H}}V_{\rm{wind}})^{1/2}$.  \cite{Martin2007} measured $n_{\rm{H}} = 0.03~$cm$^{-3}$ for the case of $o$~Ceti, a value much smaller than the canonical estimate of 1~cm$^{-3}$. Mira, itself, has a measured mass-loss rate $\dot{M} = 2.5\times 10^{-7}~M_\odot$~yr$^{-1}$ and a terminal wind speed of about 2.5~km~s$^{-1}$.  The wind parameters for V~CVn are not well-constrained, and may be greater than that of Mira.  Therefore, the bow-shock density for V~CVn may be significantly greater than that of Mira, hence would have a greater optical depth and polarization. The tail and bow shock about Mira does not produce significant polarization as the shock is so tenuous. For V~CVn, if the ISM number density is about $n_H = 1$ - $10$~cm$^{-3}$ and the dust mass-loss rate is about $10^{-8}$ - $10^{-7}~M_\odot$~yr$^{-1}$, then the optical depth of dust in the bow shock is of order unity, consistent with the optical depth necessary for the observed polarization.  

\begin{figure}[t]
\begin{center}
\includegraphics[width=0.45\textwidth]{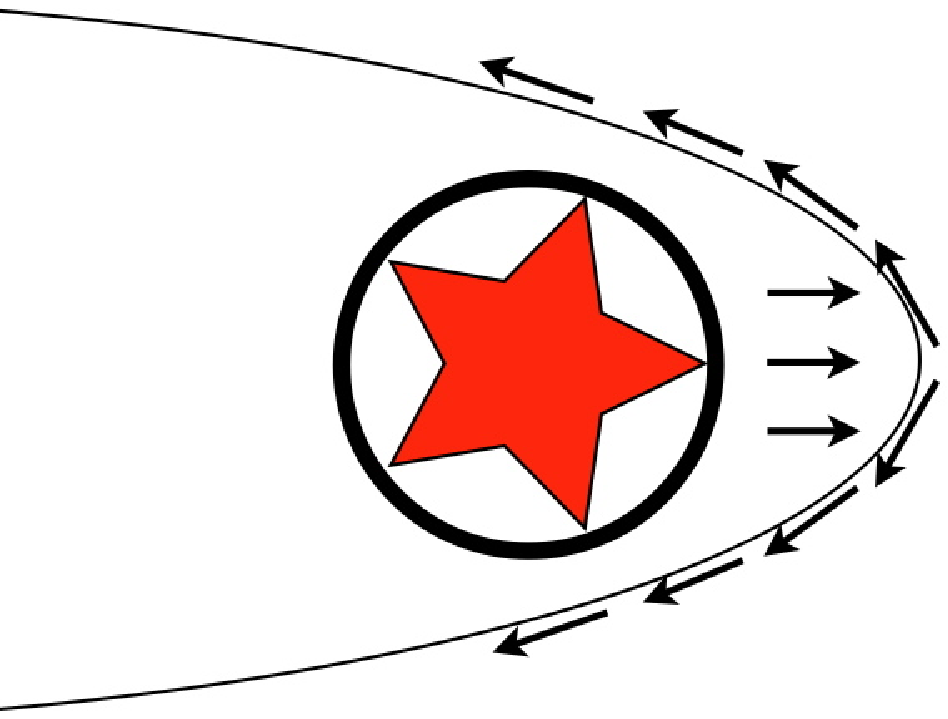}
\includegraphics[width=0.45\textwidth]{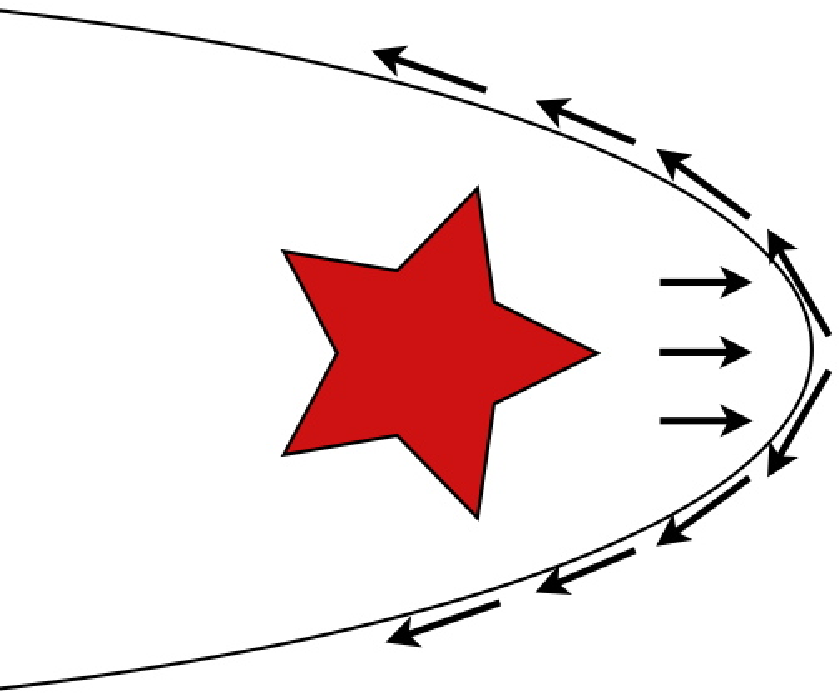}
\end{center}
\caption{ Cartoon schematically showing the interaction between the star, its pulsation-varying wind shell and the bow shock. (Top) At maximum light and minimum polarization, the dust is accelerated from the photosphere in a dense shell (denoted by the circle) and the mass-loss rate is greatest. The wind shell is denser than the bow shock and dominates the fraction of scattered light.  (Bottom) At later pulsation phases there is less radiative acceleration and dust is forming in the photosphere while the mass-loss rate is much smaller.  The shell formed at the previous flux maximum had expanded and is incorporated into the smaller density bow shock that dominates the observed polarization. The horizontal arrows represent the direction of motion of the star relative to the interstellar medium, while arrows about the stellar-wind bow shock represent the direction of motion of swept-up wind material.}\label{toon}
\end{figure}

Regarding the second challenge of the polarization variability, we argue that the polarization is modulated by variability in the stellar wind.  When V~CVn has maximum luminosity then it also has maximum pulsation velocity, and at this phase the mass-loss rate is greatest. If this shell is spherically-symmetric then the wind density is $\dot{M}/4\pi r^2 V_{\rm{wind}}$. At this pulsation phase, the shell is at location $r$, which is closest to the star, hence the wind density is greatest for any phase. The degree of polarization is a function of the asymmetric geometry of the unresolved object. Assuming a dipole-like polarigenic mechanism then $p \propto \gamma$ where $\gamma = [\int n(r,\mu)\mu^2d\mu dr]/ [\int n(r,\mu)d\mu dr]$.  The function $\gamma$ is a function of the number density as a function of distance and angle, $\mu = \cos \theta$.  Therefore, when the mass-loss rate is greatest and the wind shell is most dense then the value of $\gamma \rightarrow 1/3$ and $p $ is at its minimum. When the shell is least dense then the value of $\gamma$ is determined by the shape of the bow shock plus tail, hence most asymmetric. We present a cartoon of this scenario in Fig.~\ref{toon}.

% The prototype Mira too has a bow shock and tail (REFS)

%but no significant p, this is a density challenge  sig = sqrt(n_h * mdot), n_h is small and mdot might be orders of mag greater for hence increasing the optical depth by 1-2 orders of mag, hence p by orders of mag

% Can this explain the variable p. If wind is ejected in mass shells then yes

% In this situation, at max light most dust is ejected in symmetric outflow, because r is smallest than rho_w is greatest, hence situation is most symmetric.  The star moves and wind collides with bow shock and asymmetry dominates again.

%-----
Stellar wind bow shocks about AGB stars have been observed using the Herschel satellite \citep{Cox2011} and the Mira tail was detected by GALEX \citep{Martin2007}.  V~CVn was observed by both, however the data is insufficient to measure the presence of either a UV tail or an IR bow shock.  V~CVn is too far away to resolve structures at ultraviolet wavelengths with GALEX as it is five to ten times further away than Mira.  There is a hint of a bow shock geometry in the Herschel observations, but the resolution is limited, hence inconclusive. Higher resolution observations will improve the situation and test our hypothesis for explaining the observed polarization.   The extended asymmetric structure, observed at IR wavelengths by \cite{Kervella2014}, about L$_2$~Pup may also be interpreted as a stellar-wind bow shock. This idea is supported as the material is aligned in the north-east to south-west axis consistent with the star's direction of motion.

\section{Conclusions}
We present forty years of polarization observations from the HPOL instrument and results  \citep{Poliakova1981, Serkowski2001} previously published for the semi-regular variable red giant star V~CVn. Its measured polarization fraction at optical wavelengths varies from $1\%$ -- $6\%$ as a function of brightness, indicating a roughly inverse correlation with pulsation phase where $p$ peaks at brightness minimum.  This is consistent with the results of \cite{Kruszewski1968}.  We find that the polarization position angle has been nearly constant over the past forty years.  

We have explored a number of possible scenarios for explaining the polarimetric data such as rapid rotation, star spots, binarity, a debris disk and a stellar wind bow shock.   The first three scenarios are found to be problematic, whereas a debris disk is possible. However, the bow shock is the most likely scenario.  The advantages of a bow shock and corresponding tail is that they allow for a constant polarization position angle.   Pulsational variability can plausibly lead to opacity changes to account for the observed polarimetric variability by modulating the optical depth of the stellar wind.  A bow shock also helps to explain the overall high level of polarization at $\sim 6\%$, much greater than typically observed among other Mira and semi-regular variables, especially if the star is speeding through a denser region of the ISM than the prototype $o$~Ceti.  

%Additionally, the presence of a debris disk is also consistent with observations of white dwarfs with debris disks that are detected in around 4\% of white dwarfs \citep{Barber2012}, as well as observations of IR excesses about the central stars of planetary nebulae \citep{Bili2012}.  
%The presence of a disk in V~CVn would suggest an evolutionary scenario for understanding their presence \citep{Dong2010}, perhaps even as a precursor stage for white dwarfs with planetary companions \citep{Farihi2005}.   As such, the case of V~CVn would be the first indicator of a debris disk about an evolved giant star.  It must be noted the semi-regular variable L$_2$ Pup has also been observed to have a large polarization fraction similar to V~CVn, suggesting the potential for a second semi-regular variable star with a debris disk.

V~CVn is also  known to have an OH maser. \cite{Wolak2012} measured  circular polarization of the maser to be almost 100\% but negligible linear polarization.  This suggests that the star might be in a transition phase between being a semi-regular variable to becoming an OH/IR star \citep{Habing1996}. \cite{Lewis2004} note that OH masers in these stars are sensitive to the stellar mass-loss rate.  The presence of a maser about V~CVn may hint at the polarigenic mechanism being related to mass loss and perhaps a variable stellar wind as well as the formation of a bow shock.

  Additional modeling is badly needed; however, this will require better constraints on fundamental stellar properties.  For instance, the error on the Hipparcos parallax is about 60\% \citep{Leeuwen2007} implying limited constraints on the stellar radius.  Similarly, \cite{Tuthill1999} measured a uniform-disk angular size of about 30~mas and a (assumed Gaussian) circumstellar medium of about 15~mas using aperture masking interferometry, but limb-darkening corrections are required.  \cite{Neilson2013} noted that limb-darkening corrections for uniform-disk angular diameters could be as much as 30\% depending on the gravity and effective temperature, meaning that V~CVn has an angular diameter $30 < \theta_{\rm{LD}} < 40$~mas.    The effective temperature is about 3400~K, though this, too, is not well-constrained \citep{Magalhaes1986a}. For instance, \cite{Wasatonic1998} have suggested an even cooler value of $T_{\rm{eff}} \le 3200~$K, and \cite{Taylor2008} found that the effective temperature varied from about 2600~K to about 3200~K as a function of pulsation phase.
  
Although we do not constrain the properties of the putative stellar wind bow shock and tail, if causing the observed effects, it must be sufficiently dense to produce the observed polarization. Either the ISM density must be much greater than that about $o$~Ceti, or the mass-loss rate is much greater.  Detailed IR observations, in particular interferometric observations or IR imaging, would help to constrain bow shock properties. Similarly, if V~CVn is analogous to $o$~Ceti, then UV imaging would reveal emission in the stellar wind tail.  Initiation of a study of the closely-related star L$_2$~Pup is also important for comparative studies.

\acknowledgements The authors acknowledge support for this research through a grant from the National Science Foundation (AST-0807664).  The authors are grateful to the observers who contribute valuable observations to the AAVSO as well as to those using the HPOL instrument that made this research possible.  The authors also used the MAST database for this research.

\bibliographystyle{aa}
\bibliography{miras}
\end{document}